\documentstyle[12pt,epsf]{article}
\newcommand{\JMaP}[1]{J. Math. Phys.\ {\bf #1}\ }

\newcommand{\NC}[1]{Nuovo Cimento\ {\bf #1}\ }
\newcommand{\NP}[1]{Nucl. Phys.\ {\bf #1}\ }

\newcommand{\PL}[1]{Phys. Lett.\ {\bf #1}\ }
\newcommand{\PR}[1]{Phys. Rev.\ {\bf #1}\ }

\newcommand{\PRS}[1]{Proc. Roy. Soc. A \ {\bf #1}\ }
\newcommand{\PTP}[1]{Prog. Theor. Phys. \ {\bf #1}\ }

\newcommand{\ZP}[1]{Z. Phys.\ {\bf #1}\ }

\newcommand{\eq}{\begin{equation}}
\newcommand{\en}{\end{equation}}

\newcommand{\eqa}{\begin{eqnarray}}
\newcommand{\ena}{\end{eqnarray}}
\newcommand{\lbl}{\label}
 
\def\Dir{\nabla\kern-2ex\Big{/}}

\def\Dsl{\partial\kern-1.5ex\Big{/}}
\def\slb{{b\kern-1ex {/}}}
\def\slp{{p\kern-1ex {/}}}
\def\slq{{q\kern-1ex {/}}}
\def\slk{{k\kern-1.2ex {/}}}

\def\naturali{{\hbox{l\kern-.5ex N}}}
\def\complessi{{\hbox{l\kern-1.2ex C}}}
 
\def\pd#1{{\partial~\over\partial #1}}
 
\def\aa{\alpha}
\def\bb{\beta}

\def\dd{\delta}

\def\ee{\epsilon}

\def\gg{\gamma}

\def\LL{\Lambda}

\def\rr{\rho}

\def\ss{\sigma}

\newcommand{\ftt}[2]{{d^{#1}{#2}\over (2\pi)^{#1}}}

\def\pv{{\vec p}}
\def\qv{{\vec q}}
\def\kv{{\vec k}}

\begin{document}
\begin{flushright}
HD-THEP-97-30
\end{flushright}
\tolerance=10000
\hfuzz= 5 pt
\baselineskip=24pt

\begin{center}
\begin{Large}
{\bf The Barbieri--Remiddi solution of the bound state problem in QED}\\[2cm]
\end{Large} 
{\large Antonio Vairo}\footnote{Alexander von Humboldt Fellow}
                      \footnote{a.vairo@thphys.uni-heidelberg.de}\\
{\it Institut f\"ur Theoretische Physik, Universit\"at Heidelberg\\ 
Philosophenweg 16, D-69120 Heidelberg, FRG}
\end{center}
\vspace{1cm}

\begin{abstract}
\baselineskip=20pt
We derive the so-called Barbieri--Remiddi solution of the Bethe--Salpeter 
equation in QED in its general form and discuss its application 
to the bound state energy spectrum. 
\vspace{0.5cm} 
\end{abstract}

\setcounter{footnote}{0} 

\section{Introduction}

The Bethe--Salpeter (BS) equation \cite{Bet51} is usually considered 
the rigorous framework in which to approach the bound state problem in QED. 
The increasing precision of the experimental data concerning 
QED bound states (e. g. decay rate, energy levels, etc. for some 
recent reviews we refer the reader to \cite{Kin90}) makes more and more 
urgent to effort the evaluation of physical quantities 
by handling the BS equation with a systematic and unified formalism. 

In this paper we will focus our attention to the bound 
state energy levels in QED and will discuss the so-called 
Barbieri--Remiddi (BR) formalism. This formalism was first suggested 
for positronium \cite{Bar78,Cas78} (for a clarifying quanto-mechanical 
example see also \cite{Her95}), but has been used in recent years 
also for hydrogenic atoms \cite{Cav94}, QCD bound states \cite{Kum94} 
and scalar-scalar bound states \cite{Moe96}. The main idea 
is to solve exactly the BS equation for a suitable 
zeroth-order kernel containing the relevant binding interaction 
(i. e. the Coulomb potential) and then to perform a perturbative 
expansion in terms of the difference between the complete two-body 
kernel and the zeroth-order one. What is appealing in this approach 
is that the zeroth-order solution is completely known in analytic  
closed form. Therefore the perturbative expansion obtained in this 
way is completely self-contained and does not need to be improved 
for higher correction in the fine structure constant $\alpha$. 

In the following the BR formalism will be derived in the general 
case of muonium (i. e. different masses). This result is new 
and contains the positronium and hydrogenic case as a particular one. 
Moreover it furnishes a way to treat radiative and 
recoil corrections in the same theoretical framework, which 
seems to be very promising. 

The paper contains two main sections. In section 2 
we derive the perturbative expansion of the energy levels 
from the BS equation in the so-called Kato 
formalism. In section 3 we derive in some detail the BR 
solution for muonium. Section 4 is devoted to some comments and 
conclusions. 

\section{The Bethe--Salpeter equation}

In this section we review some basics concerning the 
Bethe--Salpeter equation in QED and set up the theoretical 
background for the next section. The main result will be the perturbative 
expansion of the energy levels of the two fermion bound state 
given by Eq. (\ref{energy}). 

\begin{figure}[htb]
\vskip 0.8truecm
\makebox[0.2truecm]{\phantom b}
\epsfxsize=14.8truecm
\epsffile{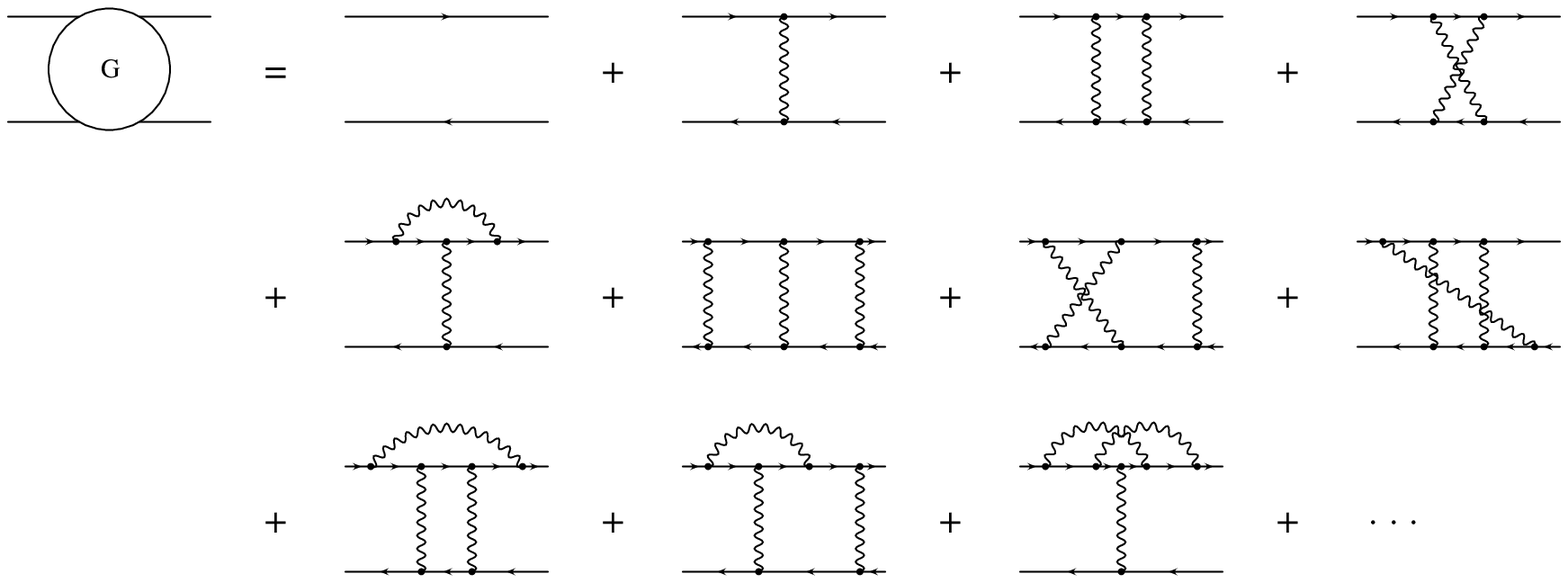}
\vskip 0.3truecm
\caption{{{\small \it Four point Green function $G$: graphs up to 
two loops.}}}
\lbl{psg}
\vskip 0.8truecm
\end{figure}

Let us consider a system of two fermions (of mass $m_1$ and $m_2$ 
and electric charge $-e$ and $Ze$ respectively) like muonium. 
The four point Green function $G$ is the sum of the Feynman graphs 
shown in Fig. \ref{psg} (notice that for a particle-antiparticle system, 
like positronium, one has to add the annihilation graphs). 
Let us define $G_0$ the two fermions free propagator: 
\eq
G_0(E;p) = S_F^{(1)}(E_1;p) S_F^{(2)}(-E_2;p)^T, \qquad\quad
S_F(E_j;p) = {i \over E_j\gg^0 + \slp -m +i\ee},
\lbl{G0}
\en
where $^{(1)}$ and $^{(2)}$ refer to the two fermion lines, 
$E_j \equiv  E ~m_j / (m_1+m_2)$, $E$ is the bound state energy   
and $p$, $q$ are the relative momenta of the outcoming and incoming 
particles in the centre-of-mass reference frame 
\footnote{
Let $p^{(1)}$ and $p^{(2)}$ be the momenta of the  
outcoming particles, and 
$$
P \equiv  p^{(1)} + p^{(2)}, \qquad\quad 
p \equiv {\mu \over m_1}p^{(1)} - {\mu \over m_2}p^{(2)},
$$
with $\mu = m_1 m_2 / (m_1 + m_2)$ the reduced mass of the two particles.
In the centre-of-mass frame $\pv^{~(1)} = - \pv^{~(2)}$  implies 
$P = (p^{(1)}_0+p^{(2)}_0,\vec 0~)$ $= (E,\vec 0~)$.
From the previous equations we obtain
\eqa
p^{(1)} &=&  p + {m_1 \over m_1+m_2}P = ( p_0 + E_1, \pv~),
\nonumber\\
p^{(2)} &=& -p + {m_2 \over m_1+m_2}P = (-p_0 + E_2,-\pv~),
\nonumber
\ena
where $E_j = E~m_j / (m_1+m_2)$ and therefore 
$$
E = E_1+E_2  .
$$
Finally, we note that in the static limit
$$
E  \rightarrow m_1+m_2, ~~~~ E_j \rightarrow m_j .
$$
In the same way we can treat the incoming momenta.}.
It is found that the Green function $G$ satisfies the equation:
\eq 
G(E;p,q)=
G_0(E;p){\Bigg[} (2\pi)^4\dd^{4}(p-q) + \int\ftt{4}{k}K(E;p,k)G(E;k,q)\Bigg].
\lbl{bs_eq}
\en
This equation (for simplicity we have neglected the spinor indices) 
is known as the Bethe--Salpeter equation \cite{Bet51}. The kernel $K$, 
describing the interaction between the two fermions, is not known 
in analytic closed form and is given by all the two particle irreducible 
graphs without external legs shown up to two loop in Fig. \ref{psk}. 
Graphically the BS equation can be represented as in Fig. \ref{psbs_g}. 

\begin{figure}[htb]
\vskip 0.8truecm
\makebox[0.2truecm]{\phantom b}
\epsfxsize=14.8truecm
\epsffile{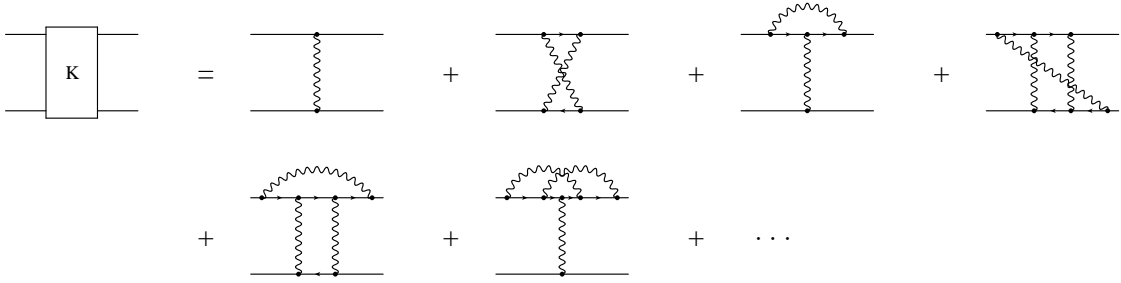}
\vskip 0.3truecm
\caption{{{\small \it The interaction kernel $K$.} }}
\lbl{psk}
\vskip 0.8truecm
\end{figure}

$G$, as a function of $E$,  has simple poles in the bound 
state energy levels $E_{n...}$ \cite{Ede53} ($n...$ is a convenient set 
of quantum numbers which classifies the levels). Since the Coulomb interaction 
is contained in $K$,  $E_{n...}$ (without mass terms) has to coincide 
with the Bohr levels at the leading order in $\aa$. 
Therefore for any ${n...}$, we can write 
\eq
G(E) = {R_{n...} \over E-E_{n...}} + \hat G_{n...}(E) \,,~~~~~~~~ 
E_{n...} = m_1 + m_2 -{\mu \over 2}
\left( {{Z\alpha} \over n} \right)^2 ~+~ \cdots ~,
\lbl{G_pol}
\en
where $R_{n...}$ is the residuum at the pole, $\hat G_{n...}$ is non singular 
in the limit $E \to  E_{n...}$ and $\mu$ is the reduced mass 
of the two particle system. From now on we will neglect the explicit 
indication of the momenta in the argument of the functions, where considered 
not strictly necessary.

\begin{figure}[htb]
\vskip 0.8truecm
\makebox[2truecm]{\phantom b}
\epsfxsize=11truecm
\epsffile{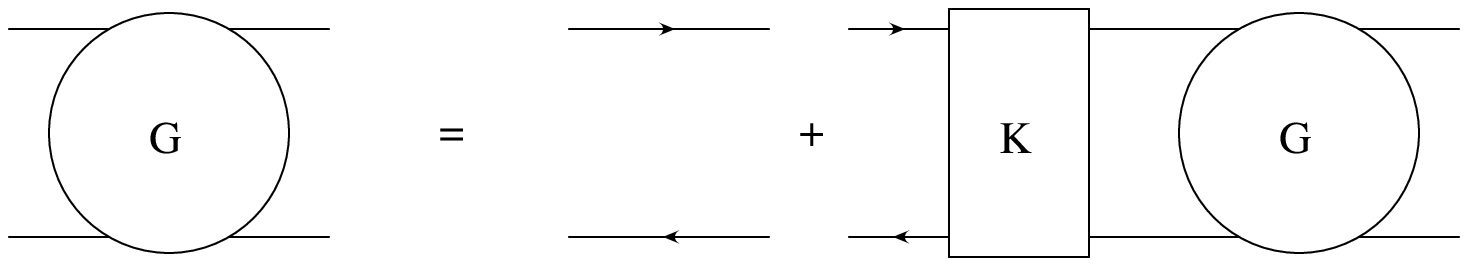}
\vskip 0.3truecm
\caption{{{\small \it Inhomogeneous BS equation: 
$G(E) = G_0(E) + G_0(E) K(E) G(E)$.}}}
\lbl{psbs_g}
\vskip 0.8truecm
\end{figure}

Inserting (\ref{G_pol}) into the BS equation and comparing the residua, 
we obtain:
\eq 
R_{n...} = G_0(E_{n...})K(E_{n...})R_{n...} , 
\lbl{bs2_eq}
\en
which is known as the homogeneous Bethe--Salpeter equation.
Moreover, from the comparison of the non singular parts in $E-E_{n...}$ 
we obtain the normalization condition \cite{Man56}:
\eq
R_{n...} = R_{n...}~\pd{E} \left( G_0^{-1} - K \right) (E_{n...})~R_{n...}~.
\lbl{norm}
\en

The BS equation (\ref{bs_eq}) is, up to now, not solvable in 
analytic closed form. Let $K_c$ be an interaction kernel satisfying 
the following two properties:
\begin{description}
\item{$i)$} ~~$K_c$ reproduces the same non relativistic limit of $K$, i.e. 
the Coulomb potential $V = -4\pi Z\aa/p^2$ times some spinorial factors;
\item{$ii)$} ~~the BS equation for $K_c$: 
\eq
G_c(E) = G_0(E) + G_0(E) K_c(E) G_c(E) ,
\lbl{Gc_eq}
\en
is analytically solvable in closed form. 
\end{description}
With these assumptions it is possible to solve the BS equation for $G$ 
at least perturbatively in terms of $\dd K \equiv K - K_c$ and to give 
a perturbative expansion for the bound state energy levels (the poles of $G$).

From the property $i)$ it follows that $G_c$ has simple poles for $E=E_n^c$. 
These poles are surely more degenerate than the poles of the 
complete Green function $G$ and give back, at the leading 
order in $\aa$, the Bohr levels:
\eq
G_c(E) = {\sum R^c_{n...} \over E-E^c_n } + \hat G^c_n(E) \,,~~~~ 
E^c_n = m_1+m_2-{\mu \over 2}\left( {{Z\alpha} \over n} \right)^2 ~+~ \cdots ~,
\lbl{Gc_pol}
\en
where $R^c_{n...}$ is the residuum at the pole 
and $\hat G^c_n$ is non singular in the limit $E \to E^c_n$. 
The sum is extended over all the degenerate states for each $n\in \naturali$.
The residuum $R^c_{n...}$ satisfies the analogous of equations 
(\ref{bs2_eq}) and (\ref{norm}):
\eqa 
R^c_{n...} &=& G_0(E^c_n)K_c(E^c_n)R^c_{n...}~, 
\lbl{gc2_eq}\\
R^c_{n...} &=& R^c_{n...} ~\pd{E} 
\left( G_0^{-1} - K_c \right) (E^c_n) ~R^c_{n...}~.
\lbl{normc}
\ena

From the definition of $\dd K$ and from (\ref{bs_eq}) and (\ref{Gc_eq}) 
we obtain the perturbative expansion of $G$ in terms of $\dd K$:
\eq
G(E) = G_c(E) + G_c(E) \dd K(E) G_c(E) 
+ G_c(E) \dd K(E) G_c(E) \dd K(E) G_c(E) + \cdots .
\lbl{expG}
\en
In order to obtain from (\ref{expG}) the perturbative expansion of the poles 
$E_{n...}$ we will use the so-called Kato perturbation theory \cite{Kat49}. 
Since the energy levels are the poles of $G$ we can write:
\eq
E_{n...} =
{\displaystyle \oint_{\kern 1pt \Gamma_n} 
{dz ~z ~{\rm Tr}\left\{ G(z)O_{n...} \right\}} \over \displaystyle 
\oint_{\kern 1pt \Gamma_n} {dz    ~{\rm Tr}\left\{ G(z)O_{n...} \right\}}},
\lbl{Egamma}
\en
where $\Gamma_n$ is a closed curve in $\complessi$ 
which contains only the poles $E_{n...}$ and $E^c_n$ of $G$ and $G_c$ 
respectively, $O_{n...}$ is an operator which does not vanish on 
$R_{n...}$ and $R^c_{n...}$ and  ${\rm Tr}$ means the trace over the spinor 
indices. A convenient choice is 
$$
O_{n...} \equiv  \pd{E} \left( G_0^{-1} - K_c \right)(E^c_n) ~ R^c_{n...} ~
\pd{E} \left( G_0^{-1} - K_c \right)(E^c_n). 
$$
Inserting (\ref{expG}) in (\ref{Egamma}) integrating in $z$ and taking 
into account (\ref{normc}), we obtain (up to order $\dd K^4$):
\eqa
E_{n...} &=& E^c_n + {1 \over D_{n...}} \left< \dd K(E^c_n) \right>_{n...} 
+ {1 \over D^2_{n...} } \left< \dd K(E^c_n) \hat G_c(E^c_n) 
\dd K(E^c_n) \right>_{n...}
\nonumber\\
&+& {1 \over D^2_{n...} } \left< \dd K(E^c_n) \right>_{n...}
\left< \pd{E} \dd K(E^c_n) \right>_{n...} 
\nonumber\\
&+& {1 \over D^3_{n...} } \left< \dd K(E^c_n) \hat G_c(E^c_n) 
\dd K(E^c_n) \hat G_c(E^c_n) \dd K(E^c_n) \right>_{n...}
\nonumber\\
&+& {1 \over D^3_{n...} } \left< \dd K(E^c_n) \hat G_c(E^c_n) 
\dd K(E^c_n) \right>_{n...} \left< \pd{E} \dd K(E^c_n) \right>_{n...} 
\nonumber\\
&+& {1 \over D^3_{n...} } \left< \pd{E} \left(\dd K \hat G_c \dd K \right) 
(E^c_n) \right>_{n...} \left< \dd K(E^c_n) \right>_{n...} 
\nonumber\\
&+& {1 \over D^3_{n...} } \left< \dd K(E^c_n) \right>_{n...}
\left[ \left< \pd{E} \dd K(E^c_n) \right>_{n...} \right]^2
\nonumber\\
&+& {1\over 2} {1 \over D^3_{n...} } 
\left[ \left< \dd K(E^c_n) \right>_{n...} \right]^2
\left< {\partial^2~ \over \partial E^2} \dd K(E^c_n) \right>_{n...} 
+ ~O(\dd K^4),
\lbl{energy}
\ena
where 
$$
\left< A \right>_{n...} \equiv {\rm Tr}\left\{ A ~R^c_{n...} \right\} = 
\int \ftt{4}{p} \int \ftt{4}{q} 
{\rm Tr}\left\{ A(p,q) ~R^c_{n...}(p,q) \right\}.
$$ 
$D_{n...}$ is the degeneracy of the $E=E_{n...}$ level and is defined to be 
\eq
D_{n...} \equiv 
\left< \pd{E} \left( G_0^{-1} - K_c \right) (E^c_n) \right>_{n...}.
\lbl{DDD}
\en

Equation (\ref{energy}) expresses the bound state energy $E_{n...}$ 
as an expansion in $\dd K$. Since $\delta K$ is the difference between 
the sum of the infinit series of Feynman graphs drawn in Fig. \ref{psk}  
and the kernel $K_c$, $\delta K$  is not known in closed form.  
Each graph of Fig. \ref{psk} contributes to (\ref{energy}) with 
a series of powers of $\aa$, because the dependence of the residuum $R^c_n$ 
on the fine structure constant (like in the well-known non-relativistic 
case where the hydrogen wave-functions depend on $\aa$). For consistency 
with $i)$ the explicit calculation must exhibit that to an increasing order in 
$\dd K$ it corresponds an increasing leading order in $\aa$ 
in the contributions to the energy levels. In this sense  
expansion (\ref{energy}) can be interpreted as a perturbative expansion in the 
fine structure constant. Once $K_c$ is explicitly given and the 
corresponding BS equation is solved (this means we have an analytic  
expression for $E^c_n$, $R^c_{n...}$ and $\hat G^c_n$) the expansion 
(\ref{energy}) allows to obtain without any ambiguity the energy levels 
up to a given order in $\aa$ for all the two-fermions bound states in QED. 
We emphasize that, in absence of an exact solution of Eq. (\ref{Gc_eq}),  
expression (\ref{energy}) could be evaluated only for an approximate choice of 
$G_c$ to improve at any increasing of the requested precision. 

\section{The Barbieri--Remiddi solution}

In this section we work out with some detail the so-called Barbieri--Remiddi 
solution of equation (\ref{Gc_eq}) (for and exhaustive description 
see \cite{Vaibo}). With this name we mean a zeroth order kernel $K_c$ 
which satisfies the previous given properties $i)$ and $ii)$ (in other words 
$K_c$ should describe correctly at the leading order in $\aa$ 
the bound state and make solvable the corresponding BS equation 
(\ref{Gc_eq})) as well as the solution of the corresponding BS equation. 
The BR solution was first given for the positronium \cite{Bar78}. 
In the following we will give the generalization of that solution 
for a bound state of two fermions with different masses, i. e. muonium. 
Once $K_c$ is given, we solve the equation for $G_c$ and 
work out the poles $E^c_n$ and the residua $R^c_{n...}$. 
At that point the perturbative expansion of the energy levels (\ref{energy}) 
will be completely defined. 

Let us define the energy projectors  
\eq
\LL_{\pm}(\pv,m_j) \equiv { E_{pj} \pm (m_j-\pv\cdot\vec\gg)\gg^0 \over
2 E_{pj}};
\en
with $E_{pj} \equiv \sqrt{\pv^{~2}+m_j^2}$. 
In terms of $\LL_{\pm}$ the free fermion propagator $S_F$ can be written as 
$$
S_F(E_j;p) = 
i \left( {\LL_{+}(\pv,m_j)\gg^0 \over p^0+E_j-E_{pj}+i\ee}
+        {\LL_{-}(\pv,m_j)\gg^0 \over p^0+E_j+E_{pj}-i\ee} \right).
$$
Moreover, we define
\eqa
\LL(\pv,\qv~) &\equiv& \left( 
{16 E_{p1}E_{p2}E_{q1}E_{q2} \over (E_{p1}+m_1)(E_{p2}+m_2)(E_{q1}+m_1)
(E_{q2}+m_2)} \right)^{1\over 2}
\lbl{LL}\\
&\times&
\left(\gg^0\LL_+(\pv,m_1){1+\gg^0 \over 2}
\LL_+(\qv,m_1)\right)^{(1)}
\left(\gg^0\LL_-(\qv,m_2){1-\gg^0 \over 2}
\LL_-(\pv,m_2)\right)^{(2)T}.
\nonumber
\ena

The zeroth order BR interaction kernel for the muonium is    
\eq
K_c(E;\pv,\qv~) \equiv i R(E;\pv~)R(E;\qv~)V(\pv-\qv~)\LL(\pv,\qv~),
\lbl{Kc}
\en
with
\eq
R(E;\pv~) = \left( {8\mu E^2 \over 
(E+E_{p1}+E_{p2})(E-E_{p1}+E_{p2})(E+E_{p1}-E_{p2})} \right)^{1\over2}.
\en
We assume (\ref{Kc}) as a definition. In the following we will verify 
that this choice satisfies the properties $i)$ and $ii)$ given in the previous 
section.  

In the static limit ($E\rightarrow m_1+m_2$, $E_j\rightarrow m_j$
and $\pv, \qv \rightarrow 0$), 
$$
K_c(E;\pv,\qv~) \rightarrow -i~V(\pv-\qv~)
\left( {1+\gg^0 \over 2} \right)^{(1)}\left( {1-\gg^0 \over 2} \right)^{(2)T},
$$
i. e. $K_c$ reproduces the Coulomb potential times some spinorial factors.

In order to verify that the choice (\ref{Kc}) makes solvable the BS 
equation (\ref{Gc_eq}) it is useful to express the Green function $G_c$ 
in terms of a new function $H_c$ \footnote{
In general $H_c$ could depend on each component of the momenta 
$p$ and $q$. The explicit calculation, however, will show   
that $H_c$ does not depend on $p_0$ and $q_0$ (see Eq. (\ref{hh_eq3})).}: 
\eq
G_c(E;p,q) \equiv G_0(E;p) + iR(E;\pv~)R(E;\qv~)H_c(E;\pv,\qv~) 
G_0(E;p)\LL(\pv,\qv~)G_0(E;q).
\lbl{Hc_def}
\en
Including (\ref{Hc_def}) and (\ref{Kc}) in (\ref{Gc_eq}), we have 
\eqa
H_c(E;\pv,\qv~) = V(\pv-\qv~) &-&i\int\ftt{4}{k}
{R(E;\kv~)^2 \over (k^0+E_1-E_{k1}+i\ee)(k^0-E_2+E_{k2}-i\ee)}
\nonumber\\
&\times& V(\pv-\kv~)H_c(E;\kv,\qv~).
\nonumber
\ena
Integrating on $k^0$ we obtain 
\eq
H_c(E;\pv,\qv~) = V(\pv-\qv~) + \int\ftt{3}{k}
{1\over E^* - k^2 /2 \mu } V(\pv-\kv~)H_c(E;\kv,\qv~), 
\lbl{eqhc2}
\en
where
$$
E^* \equiv {(E-m_1-m_2)(E-m_1+m_2)(E+m_1-m_2)(E+m_1+m_2) \over 8\mu E^2}.
$$
Equation (\ref{eqhc2}) is nothing else than the Schr\"odinger  
equation for the propagator of a non relativistic particle in 
an external Coulomb field. Therefore its solution is known. 
A way to express it is by means of the Gegenbauer polynomia 
$C^\lambda_j$ (for the definition and some properties see \cite{Ter80}) 
\cite{Sch64}:
\eq
H_c(E;\pv,\qv~) = V(\pv-\qv~) 
- {16\pi\mu(Z\alpha)^2\gamma \over (p^2+\gamma^2) (q^2+\gamma^2) } 
\sum_{j=0}^{\infty}{1\over j+1-\mu Z\alpha/\gamma} 
C^1_j\left({\hat\xi}(\pv~)\cdot{\hat\eta}(\qv~)\right),
\lbl{hh_eq3}
\en
where $\gamma\equiv \sqrt{-2\mu E^*}$. 
Substituting Eq. (\ref{hh_eq3}) in (\ref{Hc_def}) we obtain the 
explicit analytic expression of the Green function corresponding 
to the kernel $K_c$ given by (\ref{Kc}).
As we will see, once $G_c$ is given, it is straightforward to work out 
the poles $E^c_n$, the residua $R^c_{n...}$ and $\hat G^c_n$.

From (\ref{hh_eq3}) we have immediately that $G_c$ has poles in 
\eq
E^* = -{\mu\over 2}\left({{Z\alpha}\over n}\right)^2 \Rightarrow 
E = E^c_n = \sqrt{m_1^2-(\mu{Z\alpha}/n)^2} + \sqrt{m_2^2-(\mu{Z\alpha}/n)^2}
~~~ n\in \naturali .
\lbl{Enc}
\en
Notice that up to order $\aa^2$ 
$$
E^c_n \approx m_1 + m_2 - {\mu\over 2}\left({{Z\alpha}\over n}\right)^2. 
$$
The poles of $G_c$ give back the mass terms plus the Bohr levels, 
i. e. the physically correct levels up to order $\aa^2$. 
Moreover we have 
$$
E_j(E^c_n) = E^c_n{m_j \over m_1+m_2} \approx m_j 
- {m_j \over m_1+m_2}{\mu\over 2}\left({{Z\alpha}\over n}\right)^2.
$$

The residuum at the pole $E = E^c_n$, as defined in (\ref{Gc_pol}), is 
\eqa
\sum R^c_{n...}(p,q) &=& {i\over 4\mu} 
{E^c_n \over \sqrt{m_1^2-(\mu{Z\alpha}/n)^2}\sqrt{m_2^2-(\mu{Z\alpha}/n)^2}}
R(E^c_n;\pv~)R(E^c_n;\qv~)
\nonumber\\
&\times& \left( p^2+(\mu{Z\alpha}/n)^2 \right) 
\left( q^2+(\mu{Z\alpha}/n)^2 \right)
r_n(\pv,\qv~) 
\nonumber\\
&\times& G_0(E^c_n;p)\LL(\pv,\qv~)G_0(E^c_n;q) ,
\lbl{sumRc}
\ena
where 
\eq
r_n(\pv,\qv~) \equiv 64\pi n \left(\mu{Z\alpha} \over n\right)^5
{C^1_{n-1}\left({\hat\xi}(\pv~)\cdot{\hat\eta}(\qv~)\right) 
\over \left[p^2 + (\mu{Z\alpha} /n)^2\right]^2
\left[q^2 + (\mu{Z\alpha} /n)^2 \right]^2} = \sum_{l=0}^{n-1}\sum_{m=-l}^{l} 
\varphi_{nlm}(\pv~) \varphi^*_{nlm}(\qv~). 
\lbl{rr_def}
\en
$\varphi_{nlm}(\pv~) \equiv R_{nl}(p)Y_{lm}(\hat p)$ are 
the well-known hydrogen atom wave functions \cite{Bet57}. 
Moreover, we can write 
\eqa
&~&
\left(\LL_+(\pv,m_1){1+\gg^0 \over 2}
\LL_+(\qv,m_1)\gg^0\right)^{(1)}
\left(\LL_-(\qv,m_2){1-\gg^0 \over 2}
\LL_-(\pv,m_2)\gg^0\right)^{(2)T}  = 
\nonumber\\
&~&
\left(\LL_+(\pv,m_1){1+\gg^0 \over 2}
\LL_+(\qv,m_1)\gg^0\right)_{\aa\bb}
\left(\LL_-(\qv,m_2){1-\gg^0 \over 2}
\LL_-(\pv,m_2)\gg^0\right)_{\gg\dd} = 
\nonumber\\
&~&
\left( \LL_+(\pv,m_1)       \right)_{\aa\rr}
\left( {1+\gg^0 \over 2}  \right)_{\rr\ss}
\left( \LL_+(\qv,m_1)\gg^0 \right)_{\ss\bb}
\left( \LL_-(\qv,m_2)       \right)_{\gg\nu}
\left( {1-\gg^0 \over 2}  \right)_{\nu\tau}
\left( \LL_-(\pv,m_2)\gg^0 \right)_{\tau\dd}  
\nonumber\\
&~&
= {1\over 2}\sum_{Ss}
\left(\LL_+(\pv,m_1)\Gamma_{Ss}\LL_-(\pv,m_2)\gg^0\right)_{\aa\dd}
\left(\LL_-(\qv,m_2)\Gamma^{\dagger}_{Ss}
\LL_+(\qv,m_1)\gg^0\right)_{\gg\bb} ~,
\lbl{llfie}
\ena
where we have used the Fiertz identity:
$$
\left( {1+\gg^0 \over 2}  \right)_{\rr\ss} 
\left( {1-\gg^0 \over 2}  \right)_{\nu\tau} = 
{1\over 2}\sum_{Ss} \left( \Gamma^{\dagger}_{Ss} \right)_{\nu\ss}
\left( \Gamma_{Ss} \right)_{\rr\tau} ~~~ S\in{{0,1}}, ~~~s\in{{-S,...,S}} ~;
$$
with the definitions:
\eqa
\Gamma_{00} &=& {1+\gg^0 \over 2} \gg^5, ~~~ 
\Gamma_{1s}  = i{1+\gg^0 \over 2}\vec v_s\cdot \vec \gamma ~,
\nonumber\\
\Gamma^{\dagger}_{00} &=& {1-\gg^0 \over 2} \gg^5, ~~~ 
\Gamma^{\dagger}_{1s}  = i{1-\gg^0 \over 2}\vec v^{~*}_s\cdot \vec \gamma ~,
\nonumber\\
\vec v_0 &=& (0,0,1) ~~~~~~ \vec v_{\pm 1} = -{1\over \sqrt{2}}(\pm 1,i,0) ~.
\nonumber
\ena
Eqs. (\ref{rr_def}) and (\ref{llfie}) allow to identify the quantum numbers 
${n...}$ with the principal quantum number $n$, with 
the numbers $S,s$ describing the spin of the bound state and  
with the numbers $l,m$ describing the angular momentum. 
The corresponding states $R^c_{nlmSs}$ are 
\eqa
\left( R^c_{nlmSs}(p,q) \right)_{\aa\bb\gg\dd}
&=& {i\over 2\mu} 
{E^c_n \over \sqrt{m_1^2-(\mu{Z\alpha}/n)^2}\sqrt{m_2^2-(\mu{Z\alpha}/n)^2}}
R(E^c_n;\pv~)R(E^c_n;\qv~)
\nonumber\\
&\times& 
\left( { E_{p1}E_{p2}E_{q1}E_{q2} \over 
(E_{p1}+m_1)(E_{p2}+m_2)(E_{q1}+m_1)(E_{q2}+m_2) }\right)^{1\over 2}
\nonumber\\
&\times&
{\left( p^2+(\mu{Z\alpha}/n)^2 \right)
\over (p^0+E_1(E^c_n)-E_{p1}+i\ee)(p^0-E_2(E^c_n)+E_{p2}-i\ee)}
\nonumber\\
&\times&
{\left( q^2+(\mu{Z\alpha}/n)^2 \right)
\over (q^0+E_1(E^c_n)-E_{q1}+i\ee)(q^0-E_2(E^c_n)+E_{q2}-i\ee)}
\nonumber\\
&\times&
\varphi_{nlm}(\pv~) \varphi^*_{nlm}(\qv~) 
\nonumber\\
&\times&
\left(\LL_+(\pv,m_1)\Gamma_{Ss}\LL_-(\pv,m_2)\gg^0\right)_{\aa\dd}
\left(\LL_-(\qv,m_2)\Gamma^{\dagger}_{Ss}\LL_+(\qv,m_1)\gg^0\right)_{\gg\bb}.
\lbl{Rc}
\ena
These states are not degenerate (how it is possible to verify directly by 
calculating $D_{nlmSs}$). Sometimes in the literature the residua 
at the poles of the Green function are written as 
$$
\left( R^c_{nlmSs}(p,q) \right)_{\aa\bb\gg\dd}
 = \left( \psi^c_{nlmSs} (p) \right)_{\aa\dd}
   \left( \overline{\psi}^c_{nlmSs} (q) \right)_{\gg\bb};
$$
the functions $\psi^c_{nlmSs}$ and $\overline{\psi}^c_{nlmSs}$ 
are called the BR wave functions of the bound state:
\eqa
\left( \psi^c_{nlmSs}(p) \right)_{\aa\dd}
&=& \left( {i\over 2\mu} 
{E^c_n \over \sqrt{m_1^2-(\mu{Z\alpha}/n)^2}
\sqrt{m_2^2-(\mu{Z\alpha}/n)^2}} \right)^{1\over 2}
R(E^c_n;\pv~)
\nonumber\\
&\times& 
\left( { E_{p1}E_{p2} \over 
(E_{p1}+m_1)(E_{p2}+m_2)}\right)^{1\over 2}
\nonumber\\
&\times&
{\left( p^2+(\mu{Z\alpha}/n)^2 \right) 
\over (p^0+E_1(E^c_n)-E_{p1}+i\ee)(p^0-E_2(E^c_n)+E_{p2}-i\ee)}
\nonumber\\
&\times&
\varphi_{nlm}(\pv~)
\left(\LL_+(\pv,m_1)\Gamma_{Ss}\LL_-(\pv,m_2)\gg^0\right)_{\aa\dd} ~,
\lbl{psic}\\
\overline{\psi}^c_{nlmSs} &=& \gg^0 \left( \psi^c_{nlmSs} \right) ^{\dagger} 
\gg^0 ~.
\lbl{psicbar}
\ena
Actually there are no reasons to introduce 
the bound state wave functions: from the formalism developed in the 
previous section it is clear that all the physical quantities 
can be expressed in terms of residua.

Finally, we obtain $\hat G^c_n$ subtracting from $G_c$ the 
singular part: 
\eqa
\hat G^c_n(E^c_n;p,q) &=& (2\pi)^4\dd^{4}(p-q)G_0(E^c_n;p) + 
G_0(E^c_n;p)K_c(E^c_n;\pv,\qv~)G_0(E^c_n;q)
\nonumber\\
&+& \hat H_c(E^c_n;p,q) ,
\lbl{hatGc}
\ena
where $\hat H_c$ takes into account all the contributions non singular 
in $E = E^c_n$ which come from the second term in (\ref{hh_eq3}). 
For an explicit expression of $\hat H_c$ in the positronium 
case we refer the reader to \cite{Buc80b}.

\section{Conclusions}

From the above given expressions we can recover some interesting 
limiting cases. 

Putting $Z=1$ and $m_1 =m_2 \equiv m$ ($\mu = m / 2$ and 
$E_1 = E_2 = E /2$) the above given zeroth-order solution of the BS equation 
reduces to the original BR solution given in \cite{Bar78} for positronium. 
The main difference with the muonium case is that for positronium 
also annihilation graphs contribute to the interaction 
kernel $K$. In the literature the singlet state ($S =0$) is usually referred 
as {\it parapositronium} and the triplet state ($S =1$) as 
{\it orthopositronium}. Some applications can be found in 
\cite{Vaibo,Buc80b,Buc80a,Hil89,Vaied}. 

Taking the limit of one particle mass to infinity the case of a particle 
in an external Coulomb field is recovered. 
In this case (e. g. $m_2 \rightarrow \infty$ ) $\mu = m_1 \equiv m$, 
the difference ~$E_2-m_2$ ~is finite and the bound state energy $E$ 
is given by
$$
E = {\lim_{m_2 \rightarrow \infty}} (E_1+E_2-m_2).
$$
This case has been extensively studied in \cite{Cav94,Vaibo,Rem84} 
for the evaluation of the pure radiative corrections to the 
energy levels of hydrogenic atoms.  

As a conclusive remark, we stress that all these systems 
and muonium can be studied now in the same framework. 
In particular the formalism provides a powerful tool in dealing 
simultaneously with radiative, recoil and radiative-recoil corrections. 
Extremely interesting seems also to be the study of the infinit mass limit 
of one particle in the energy expansion (\ref{energy}) evaluated on 
muonium states. This should eventually clarify how this limit works 
in a purely off mass-shell context.

\vfill\eject

\end{document}